# Design and Optimization of a Pole-less 0.2 T C-Type MRI Magnet

Ivan Etoku Oiye[1], Ajay Sharma[1], and Sairam Geethanath[1*]

[1]Laboratory for Accessible MRI, Division of Cancer Imaging Research, Radiology and Radiological Sciences, Johns Hopkins University School of Medicine, 600 N. Wolfe Street, Baltimore, 21287, MD, United States

**Low-field MRI is increasingly considered accessible for imaging owing to its lower cost, simpler infrastructure requirements, and potential for mobile and point-of-care deployment. A central challenge is achieving clinically useful field strength and homogeneity while keeping the magnet lightweight and maintaining patient accessibility. This work presents the design and magnetostatic simulation of a pole-less, 0.2T, C-type bipolar magnet comprising two cylindrical N52 permanent magnets and four concentric rings that replace traditional pole pieces to enhance field homogeneity and reduce weight in bipolar magnet designs. Geometric parameters, including each magnet ring thickness, height, angular anchorage, spacing between magnets, and the magnets' vertical offset relative to the horizontal yokes, were manually investigated to improve magnetic field homogeneity in a 20 cm DSV. Simulations were performed in CST Studio Suite, yielding a peak field of 0.2T, with a peak inhomogeneity of 1.43 mT across the 20 cm DSV and a total weight of 590 kg. A pole piece design with comparable dimensions, used as a benchmark for inhomogeneity and weight, was designed and simulated. It yielded a peak field of 0.15T and a weight of 890 kg, with a 0.7 mT inhomogeneity over a 20cm DSV. This study demonstrates the feasibility of replacing the traditional pole pieces with magnet rings to reduce weight while enhancing patient access with the C-magnet structure in yoked MRI systems.**

*Index Terms* — **Magnetic resonance imaging, Low-field, Pole-less.**

## I. INTRODUCTION

Recent progress in low-field MRI technology has focused on developing lightweight, portable systems aimed at expanding diagnostic accessibility beyond conventional hospital settings[1]. Halbach arrays [2], ring-pair[3] and bipolar magnet [4] designs have emerged as key architectures for such systems. Halbach and ring-pair configurations exhibit desirable attributes, including compact geometry, low weight, and high magnetic field efficiency [5]. However, these designs typically suffer from low signal-to-noise ratio (SNR), limited patient access, high sensitivity to electromagnetic interference, and extended acquisition times [6,7]. In contrast, bipolar magnets offer higher field strength, a larger DSV, and superior field homogeneity [4]. However, they are inherently heavy due to the substantial yokes and pole pieces. Wei et al. addressed this limitation by replacing the pole pieces with a ring of shim magnets in a 0.2T H-type magnet weighing 654 kg, yielding 1.3 mT inhomogeneity across a 20 cm DSV [8]. Building upon these advancements, the present work introduces a novel pole-less, 0.2T, C-type bipolar magnet design with two concentric ring magnets that lowers system weight (600 kg) and enhances patient accessibility compared to the H-type while maintaining low magnetic field inhomogeneity (1.43 mT) over a 20 cm DSV. For benchmarking purposes, a conventional pole-piece configuration with comparable dimensions was also designed and simulated to establish reference values for field homogeneity and weight.

## II. METHODS

### A. Theory

Pole pieces and concentric permanent-magnet rings shape magnetic fields through fundamentally different mechanisms discussed below, leading to different consequences for field homogeneity, mass, and saturation behavior. By exploiting these differences, a pole-less magnet architecture can be used to advance low-field bipolar MRI systems.

*Mechanisms of field shaping:* High-permeability pole pieces provide a low-reluctance path that spreads flux from the permanent magnets and smooths the magnetostatic potential across the DSV [9]. By tailoring the pole surface into different shapes, such as conical, spherical, or optimally contoured profiles, and applying shims on the pole face [13][10], higher-order spatial field variations within the imaging region can be reduced [11]. Concentric permanent-magnet rings, in contrast, shape the net field by superposing the fields produced by individual rings. Each ring's field can be decomposed into axial and radial harmonics, and by selecting ring radii, widths, axial positions, and remanence, the ensemble can be configured to synthesize the desired low-order field profile within the DSV [12]. This modular harmonic-synthesis approach offers systematic control of field components without relying on a single large ferromagnetic element.

*Saturation behavior:* Pole pieces channel flux through limited cross sections and can locally saturate at high flux densities[13], which constrains the attainable central field unless pole mass or cross section is increased.

Conversely, multi-ring magnet configurations distribute flux shaping among permanent magnets [14] rather than forcing large flux through small soft-iron regions. This distribution mitigates the risk of local saturation and reduces overall performance dependence on the nonlinear B-H response of iron.

*Mass and homogeneity:* While high-permeability pole pieces offer efficient flux guidance and superior homogeneity [4], they incur very large masses. For example, Liu et al. reported an H-shaped permanent-magnet assembly weighing 750 kg that produced 55 mT with a 0.01375 mT inhomogeneity over a 24 cm DSV [15], and Zhao et al. constructed an H-type scanner weighing 1300 kg with 48 mT peak field and 0.0096 mT inhomogeneity over a 40 cm DSV [16]. By contrast, Wei et al. demonstrated a pole-less scanner weighing 654 kg exhibiting a 1.3 mT inhomogeneity over a 20 cm DSV [8]. Replacing pole pieces with concentric shim rings in bipolar magnet assemblies can yield higher peak fields at the expense of a moderate increase in inhomogeneity, which can be corrected with small passive magnets and dedicated corrective coils at selected locations, avoiding the large mass penalty of full pole pieces. This trade-off indicates that transferring the field-shaping role from bulk iron to permanent-magnet rings is a viable approach to advancing bipolar low-field MRI magnets.

### B. Geometric model

The design comprises two cylindrical N52 permanent magnets, each 5.08 cm thick and 20.3 cm in diameter, together with four concentric rings of varying thicknesses and heights mounted at prescribed angular orientations and separations. Steel yokes were integrated to enhance mechanical rigidity, yielding a C-shaped configuration with a 28.2 cm air gap and a DSV of 20 cm.

Fig. 1 depicts the geometric design parameters investigated, including the thickness and height of each shim ring, inter-magnet spacing, angular anchorage of the rings, and the vertical offset of the magnets relative to the horizontal yokes. A pole-piece configuration with dimensions matched to the pole-less design shown in Fig. 2 is used as the gold standard for homogeneity and as a benchmark for system weight.

### C. Simulation

Magnetostatic simulations of both the pole-less and pole-piece configurations were performed in CST Studio Suite (Dassault Systèmes, Inc., USA) to assess the magnetic-field inhomogeneity within a 20 cm DSV. Design variables were tuned to minimize field inhomogeneity while preserving a minimum air gap of 28.2 cm for patient access, a minimum peak field of 0.2T, and a maximum system weight of 600 kg to enhance portability. The parameter adjustments were performed manually to evaluate the individual influence of each design variable on system mass and field homogeneity within the 20 cm DSV.

### III. RESULTS

Fig. 3 shows the simulation results for the magnetic field distribution of the 3D pole-less magnet model in the 20 cm DSV, with a central magnetic field of 0.2T and a 1.43 mT inhomogeneity. The pole piece design used as the reference design produced a 0.15T central magnetic field in a simulation with 0.7 mT inhomogeneity in a 20cm DSV.

### A. Impact of parameter variations on field inhomogeneity in the pole-less design

Fig. 4 illustrates the results obtained from the manual parametric variations.

*The angle of ring anchorage* changes the azimuthal distribution of magnetization and, therefore, the amplitudes and phases of the spatial harmonics within the DSV. Certain angles reduce dominant aberrant harmonics and yield a local minimum in inhomogeneity, while other angles amplify those same harmonics and degrade field uniformity. This sensitivity arises because ring angle alters how the ring contributions superpose with the primary field from the cylindrical magnets and with each other, enabling partial cancellation of low-order errors when properly phased. The inner ring exhibits low sensitivity with a minimum field inhomogeneity at 0°. The outer ring is more sensitive and yields a minimum inhomogeneity at 0° anchorage.

*Inter-magnet spacing:* Reducing spacing increases the local field strength and the contribution of higher spatial harmonics, which can either amplify inhomogeneity or enable targeted cancellation of dominant low-order errors when positioned appropriately. Increasing spacing produces a broader, lower-amplitude corrective field with reduced capacity to compensate for localized deviations. The optimal inter-ring spacing was 0.72 cm, and inner ring to center magnet was 0.2 cm

*Ring thickness*: Ring thickness modifies the radial distribution of magnetization and therefore alters the amplitudes of lower-order harmonics that dominate inhomogeneity across the DSV. Changes in inner-ring thickness produced only small variations in inhomogeneity, indicating that the inner ring primarily provides a baseline shaping of the central field, whereas variations in outer-ring thickness produced a more pronounced, non-monotonic effect. Inhomogeneity decreases as the outer thickness approaches an optimal value of 7.06cm and the inner ring height of 7.76 cm.

The vertical offset of the magnets modifies the coupling between the cylindrical magnets, shim rings, and the yoke, which in turn modifies the balance of low-order spatial harmonics within the DSV. The outer-ring offset shows comparatively weak sensitivity as compared to the inner ring and center magnet within the tested range, suggesting that its contribution is dominated by broader field shaping rather than fine harmonic trimming.

The height of ring magnets changes the axial extent of the corrective field and, consequently, the spatial harmonic content within the DSV. The inner-ring height exhibits particularly strong sensitivity, indicating that axial coverage of the inner ring plays a key role in shaping the dominant low-order field terms in the DSV; the outer-ring height shows a similar but less sensitive response. Ring height was treated as a primary optimization variable and tuned jointly with ring thickness, angular anchorage, and spacing to achieve the desired harmonic cancellation and minimize inhomogeneity over the target DSV.

### B. Impact of parameter variations on system weight in the pole-less design

Angle of ring anchorage: Increasing the angle of anchorage increases the volume of ring supporting structure on the horizontal yoke, leading to an increase in system weight.

Inter-magnet spacing: Increasing it significantly increases the system mass and alters the volume of the horizontal yoke and the magnet ring radius, thereby affecting the system weight.

Effect of ring thickness on weight: Increasing the thickness of the magnet rings produces a nearly linear increase in total system mass for a fixed ring height and radius, since additional magnetic material contributes directly to the assembly weight, the geometric proportions of the horizontal yoke increase proportionally, while the support structures remain essentially unchanged. The thickness should be increased only to the extent needed to meet field and homogeneity targets.

Height of the rings: Increasing the height of the magnet rings increases the permanent magnet volume proportionally for fixed radii and thickness, and therefore produces an approximately linear rise in total system weight.

Vertical offset of the magnets: The vertical offset is a positional design parameter that translates the magnets relative to the horizontal yokes and the DSV without changing their physical dimensions. Increasing this offset (moving the magnets farther from the DSV toward the yokes) reduces the amount of iron required in the vertical yoke, thereby decreasing overall system weight; decreasing the offset produces the opposite trend.

### C. Yoke Surface Effects on Magnetic Flux Pattern and Uniformity

Surface profiling of the horizontal yokes provides an effective passive shimming mechanism for improving field homogeneity in bipolar MRI magnets by reshaping the magnetic flux-return path and modifying the magnetostatic boundary conditions at the air-steel interface. The yokes form a high-permeability boundary, therefore, variations in local thickness or surface curvature change the effective magnetic reluctance and consequently redistribute the $B_0$ field within the DSV.
Central surface features, such as cylindrical cuts (denoted as (C) in Fig. 5) or ring boss extrusions (denoted as (B) in Fig. 5) near the yoke midline, primarily influence the on-axis field and its axial decay by concentrating flux near the centerline. This geometric bias modifies the balance of axial harmonic terms and can be used to reduce center-to-edge variation along the DSV axis. In contrast, conical cuts (denoted as (A) in Fig. 5) preferentially influence off-axis flux paths by redistributing flux at specific radii, making them effective for trimming radial inhomogeneity components that dominate toward the DSV periphery. Consequently, combined profiles (A+B) and (B+C) increase the available degrees of freedom for harmonic shaping and can simultaneously suppress both axial and radial nonuniformity terms if feature dimensions and placement are appropriately tuned.

The homogeneity benefit of yoke surface profiling is strongly geometry-dependent and exhibits coupled interactions with other magnet design variables, such as the thickness and height of each shim ring, inter-magnet spacing, angular anchorage of the rings, and the vertical offset of the magnets relative to the horizontal yokes. Minor variations in feature depth, radius, or axial placement can produce appreciable changes in field homogeneity; accordingly, these parameters should be treated as optimization variables rather than fixed mechanical details. Surface profiling also entails tradeoffs: protrusions increase mass, whereas recesses reduce mass but can reduce flux-carrying capacity. Therefore, yoke surface profiles should be co-optimized with the overall magnetic circuit to improve homogeneity within the target DSV without degrading peak field strength, structural integrity, or saturation margin.

Fig .5 illustrates the set of unoptimized profiles incorporated into the horizontal yoke and their corresponding effects on the magnetic field within the DSV.

## IV. DISCUSSION

The pole-less C-type magnet demonstrates that replacing heavy soft-iron pole pieces with concentric permanent-magnet rings is a viable route to boost field strength in low field MRI systems while substantially reducing system mass and improving patient access. Magnetostatic simulations indicate that the pole-less design achieves a central field of 0.2T with an inhomogeneity of 1.43mT over a 20 cm DSV and a total mass of 590 kg. Alternatively, the pole-piece reference geometry produced a lower central field (150 mT) but with a superior homogeneity (0.7 mT inhomogeneity) at a substantially higher mass (890 kg). These results show a tradeoff of shifting flux-shaping from bulk soft iron to permanent-magnet rings, reducing system weight and increasing achievable peak field at the expense of somewhat increased field inhomogeneity. The observed increase in static inhomogeneity for the pole-less design arises from the discrete ring geometry and the lack of a continuous high-permeability path to smooth flux lines. The parameter study demonstrated that ring thickness, height, angle of anchorage, spacing, and vertical offset relative to the yokes each measurably influence low-order and higher-order field terms within the DSV. Passive shimming with small magnets and active shimming through dedicated low-order coils at selected specific locations can be integrated to target dominant low-order errors without reclaiming the large mass penalty of full pole pieces.

### A. Patient accessibility

The C-shaped bipolar geometry with a 28.2 cm air gap and a 20 cm DSV opens one side of the magnet, enabling lateral access to the patient that is limited to closed-bore or yoke-dominated H-magnets. This open configuration facilitates positioning of anxious, pediatric, or larger patients and supports interventional workflows by improving line-of-sight and physical access for tools, staff, and monitoring equipment around the region of interest.

### B. Safety in assembly

Owing to their strong attraction to the steel yokes compared with mutual magnet-magnet interactions, the permanent magnets are effectively self-retained, reducing the need for additional support structures. The optimized configuration produced a peak force of 3100 N on the yokes; consequently, a topology optimization of the yokes was performed to accommodate these forces, followed by displacement and stress analyses to verify structural safety. For improved manufacturability, lower cost, and safer assembly, the solid ring magnets were segmented into multiple arc magnets, thereby reducing peak handling loads and mitigating massive attractive forces during assembly.

### C. Future work

Future work will replace the present manual parametric tuning with an automated optimization framework that simultaneously targets central field strength, inhomogeneity within the 20 cm DSV, and total mass. In parallel, both passive and active shimming strategies will be investigated to mitigate the residual inhomogeneity of the pole-less configuration. Passive shimming will include optimized placement of small permanent-magnet, whereas active shimming will evaluate low-order shim coil sets. The combined effect of automated design optimization and hybrid shimming will be quantified by using field maps to determine the achievable corrected homogeneity.

## V. CONCLUSION

The pole-less C-type bipolar magnet is a promising architecture for lightweight, accessible low-field MRI. By replacing conventional pole pieces with concentric permanent-magnet rings for flux shaping, the design achieves a 0.2T central field while improving patient access and reducing total mass, at the cost of a modest increase in native field inhomogeneity relative to pole-piece systems.

## Acknowledgement

This research was funded by the Johns Hopkins Provost's DELTA Award (PI: Geethanath) and Medirays Healthcare Pvt. Ltd (PI: Geethanath).

Figures

**Fig. 1**

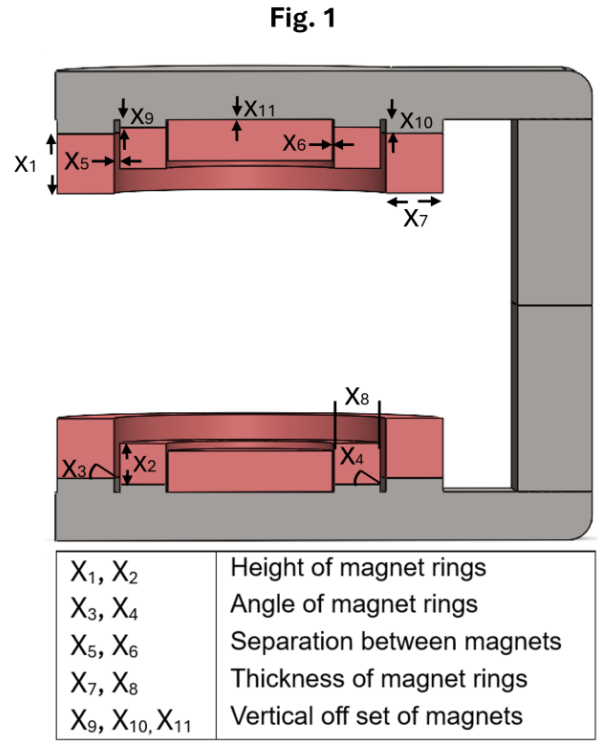

| $X_1, X_2$ | Height of magnet rings |
| $X_3, X_4$ | Angle of magnet rings |
| $X_5, X_6$ | Separation between magnets |
| $X_7, X_8$ | Thickness of magnet rings |
| $X_9, X_{10}, X_{11}$ | Vertical off set of magnets |

Fig. 1: Geometric design variables, including the thickness and height of each shim ring, inter-magnet spacing, angular anchoring of each ring, and vertical offset of the magnets relative to the horizontal yokes.

**Fig. 2**

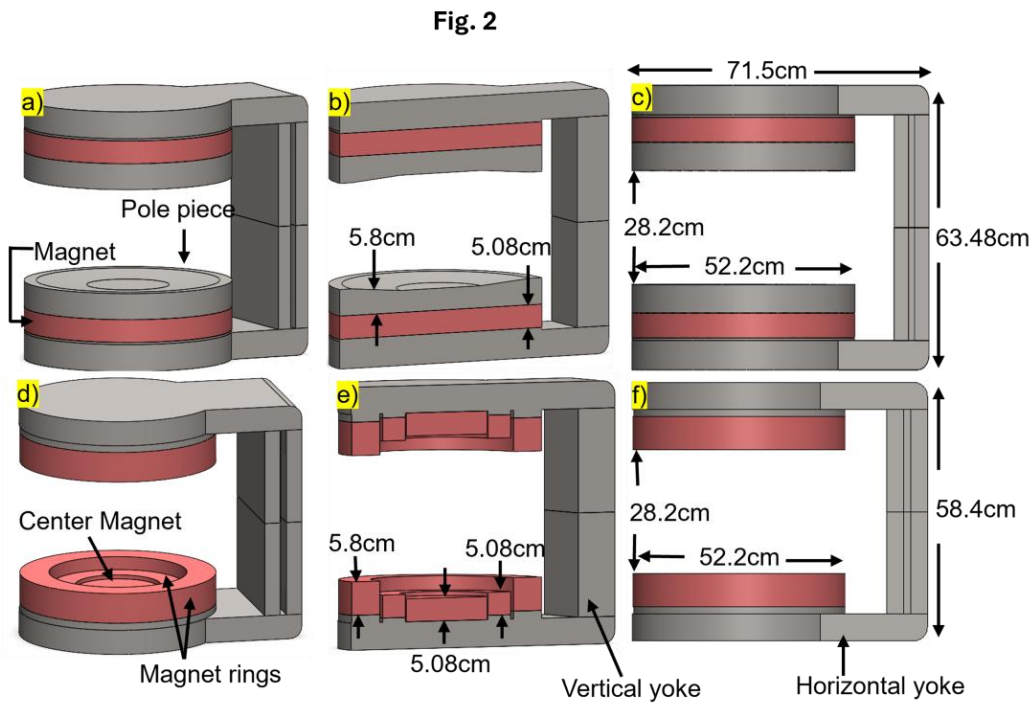

Fig. 2: Magnet configurations: a-c) CAD models of C-type magnet configuration with pole pieces; d-f) CAD models of pole-less C-type magnet configuration.

**Fig. 3**

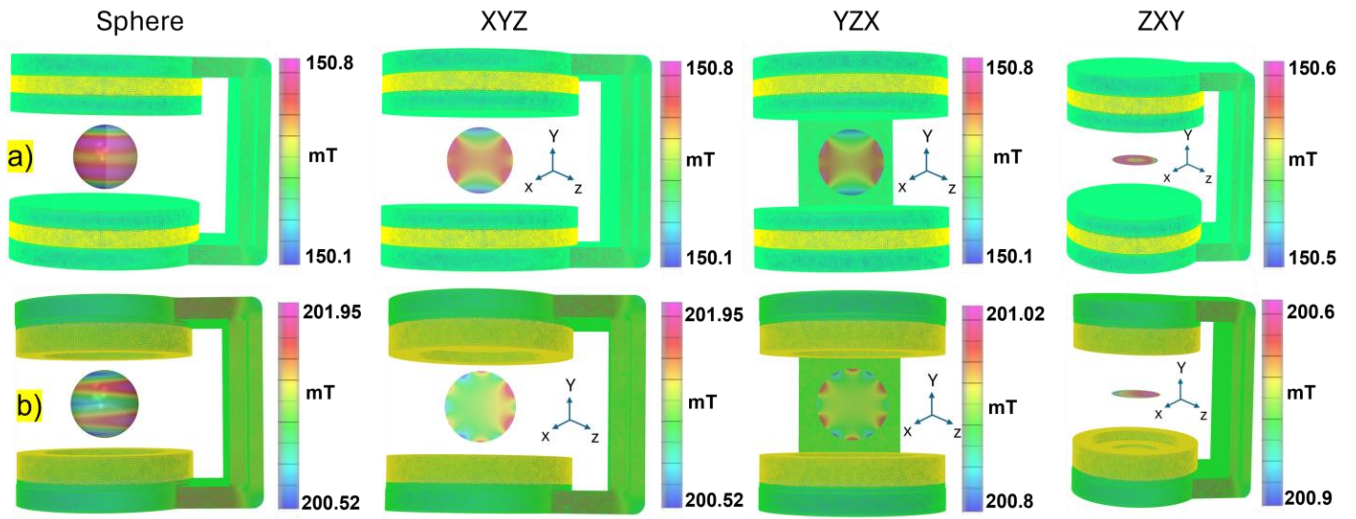

Fig. 3: Magnetic field map in 20cm DSV: a) magnetic field map of the pole pieces configuration outputting 0.7 mT inhomogeneity; b) magnetic field map. of the pole-less C-type magnet producing 1.43 mT inhomogeneity.

**Fig. 4**

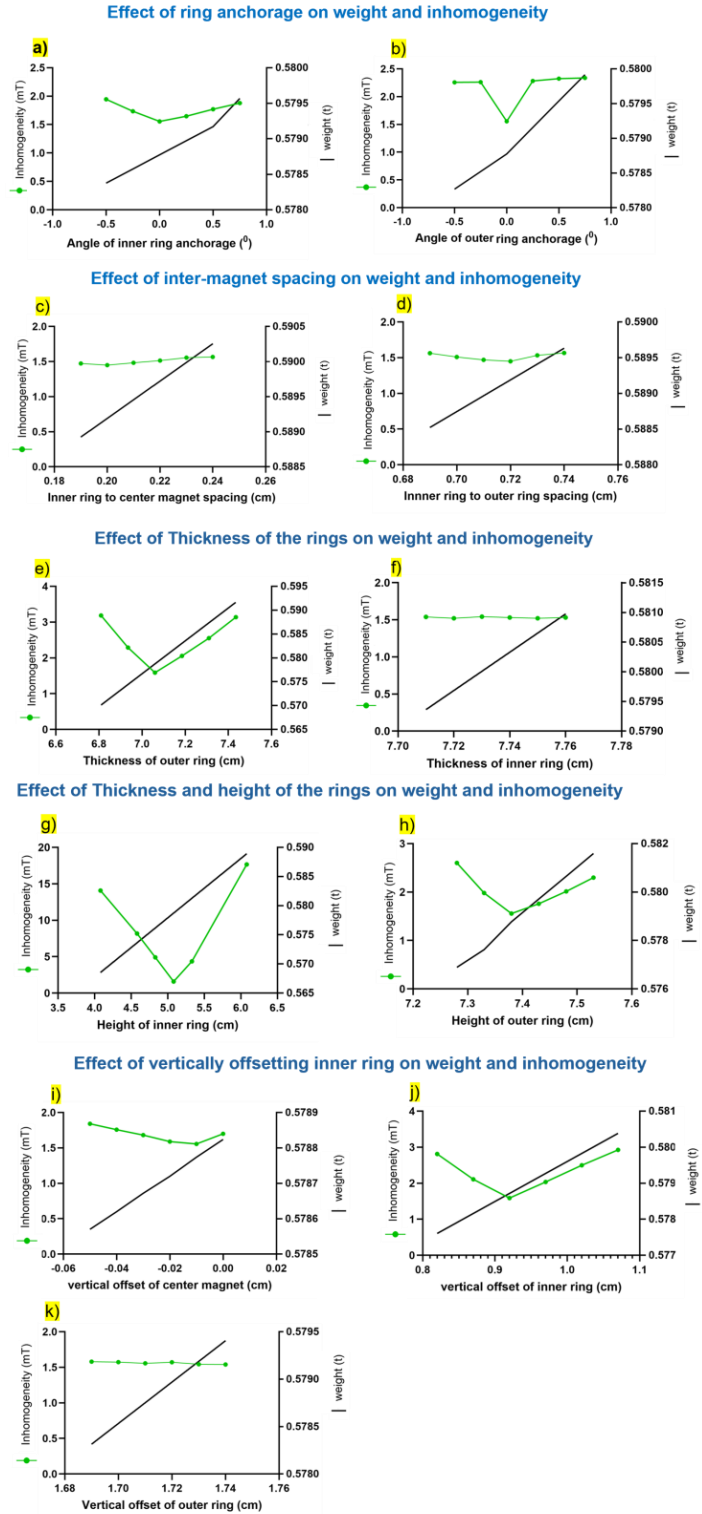

Fig. 4 illustrates the influence of geometric parameters on system mass and field inhomogeneity: a)-b) inner- and outer-ring anchorage angle, c)-d) inter-magnet spacing, e)-f) ring thickness, g)-h) ring height, and i)-k) magnet vertical offset.

**Fig. 5**

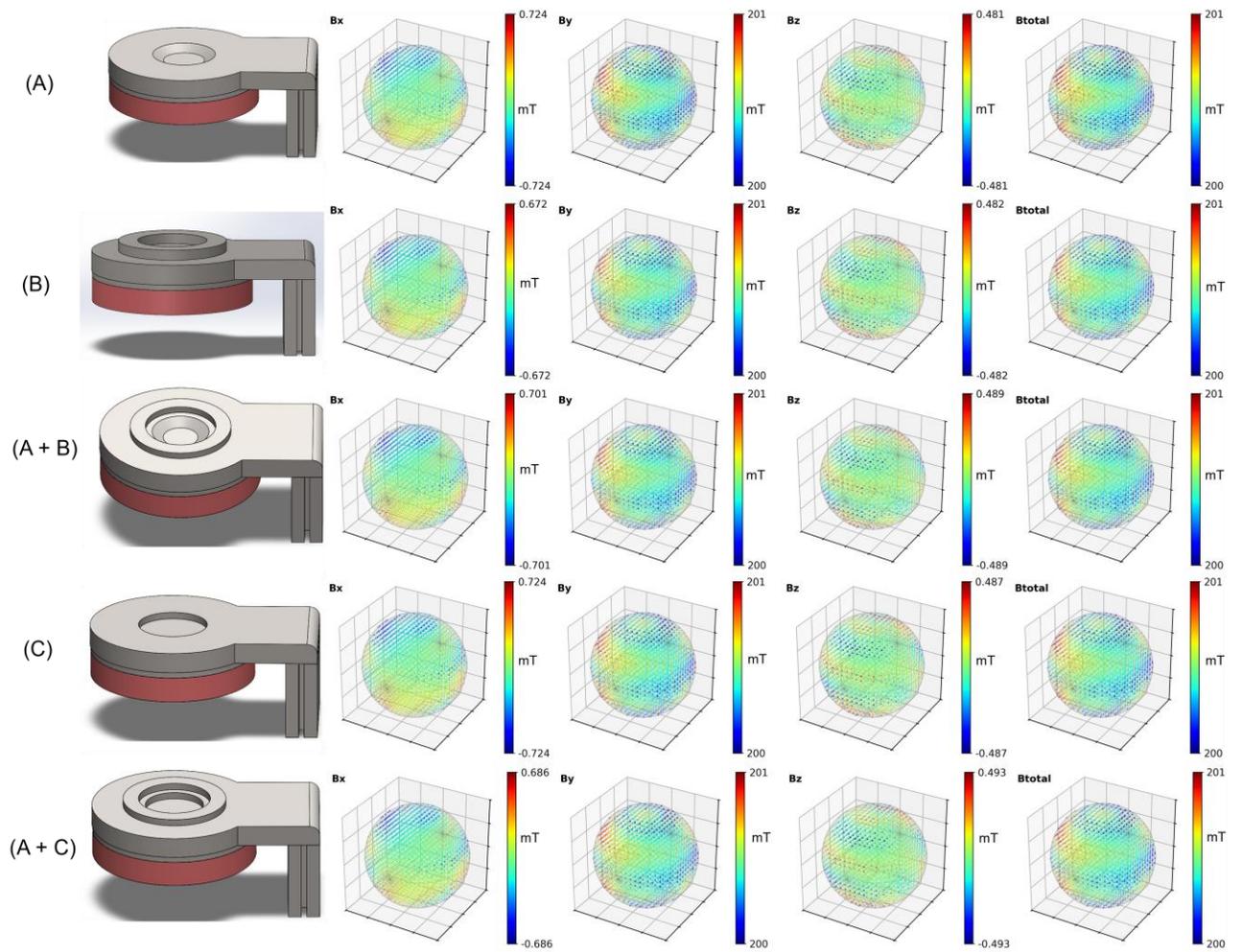

Fig. 5 presents the horizontal-yoke surface profiles and the corresponding homogeneity map within the DSV: (A) conical cut, (B) ring extrusion, (A+B) combined conical cut and ring extrusion, (C) cylindrical cut, and (B+C) combined ring extrusion and cylindrical cut.